\documentclass[nofootinbib,aps,reprint,superscriptaddress]{revtex4}
\usepackage{dcolumn}
\usepackage{bm}
\usepackage{pdfpages}
\usepackage{amsmath}    % need for subequations
\usepackage{amsfonts} % note how statements can be commented out
\usepackage{amssymb}
\usepackage{mathptmx}      % use Times fonts if available on your TeX system
\usepackage{graphicx}   % for figures
\usepackage{mathtools}
\usepackage{graphics}
\usepackage{epsfig}
\usepackage{tabularx}   % fit table to page width
\usepackage{multirow}
\usepackage{bibentry}
\usepackage{braket}
\usepackage{algorithm}
\usepackage{algorithmic}
\usepackage{listings}
\usepackage{setspace}
\usepackage{fancyhdr}
\usepackage{slashed}
\usepackage{color}
\usepackage{xfrac}
\usepackage[pdftex, pdfborder= 0 0 0, citecolor=blue, urlcolor=blue, linkcolor=blue, colorlinks=true, bookmarksopen=true]{hyperref}
\def\Journal#1#2#3#4{{#1} {\bf#2}, #3 (#4)}

\def\NPB{{\rm Nucl. Phys.} B}
\def\PLB{{\rm Phys. Lett.}  B}
\def\PRL{\rm Phys. Rev. Lett.}
\def\PRD{{\rm Phys. Rev.} D}
\def\PRC{{\rm Phys. Rev.} C}
\def\ZPC{{\rm Z. Phys.} C}
\def\JPG{{\rm J. Phys.} G}
\def\EPJC{{\rm Eur. Phys. J.} C}

\def\ep{\epsilon}
\def\vep{\varepsilon}
\def\la{\langle}
\def\ra{\rangle}

\def\be{\begin{equation}}
\def\ee{\end{equation}}
\def\bea{\begin{eqnarray}}
\def\eea{\end{eqnarray}}
\begin{document}
%\begin{frontmatter}
%\vspace{2.in}
%\def\correspondingauthor{\footnote{Corresponding author: homyoung@knu.ac.kr}}
\title{Systematic twist expansion of $(\eta_c,\eta_b)\to\gamma^*\gamma$ transition form factors in light-front quark model}
\author{  Hui-Young Ryu}
\affiliation{\em Department of Physics, Pusan National University,
     Pusan, Korea 46241}
     
\author{ Ho-Meoyng Choi}
\email{homyoung@knu.ac.kr}
\affiliation{\em Department of Physics, Teachers College, Kyungpook National University,
     Daegu, Korea 41566}
     
 \author{Chueng-Ryong Ji}
\affiliation{\em Department of Physics, North Carolina State University,
Raleigh, NC 27695-8202} 
\begin{abstract}
The light-front quark model analysis of the meson-photon transition form factor $F_{P\gamma} (Q^2)$ amenable 
both for the spacelike region ($Q^2 >0$) and the timelike region ($Q^2 <0$) provides a systematic twist expansion
of $Q^2 F_{P\gamma} (Q^2)$ for the high $|Q^2|$ region. Investigating $F_{P\gamma} (Q^2) (P = \eta_c,\eta_b)$
for the entire kinematic regions of $Q^2$, we examine the twist-2 and twist-3 distribution 
amplitudes of $(\eta_c,\eta_b)$ mesons in the light-front quark model and quantify their contributions
to  $Q^2 F_{(\eta_c,\eta_b)\gamma}(Q^2)$. 
Our numerical results for the normalized transition form factor $F_{(\eta_c,\eta_b)\gamma}(Q^2)/F_{(\eta_c,\eta_b)\gamma}(0)$ and the decay width
$\Gamma_{(\eta_c,\eta_b)\to\gamma\gamma}$ are compared with the available data checking the sensitivity of our model to the variation of
the  constituent quark masses. 
%The  scaling behaviors $|Q^2F_{(\eta_c,\eta_b)\gamma}(Q^2)|$ for large $|Q^2|$  are obtained.
\end{abstract}
%\begin{keyword}
%Transition form factors, Heavy quarkonia, Light-front quark model
%\end{keyword}
\maketitle
%\date{}
%\end{frontmatter}

\section{Introduction}
\label{sec:I}
The pseudoscalar meson ($P$) production processes via the two-photon collision, $\gamma^*\gamma\to P$,
involve only one transition form factor (TFF) $F_{P\gamma}(Q^2)$, where $q^2=-Q^2$ is
the squared momentum transfer of the virtual photon. This meson-photon transition is well known to be the simplest 
exclusive process
in testing the quantum chromodynamics (QCD) and understanding the structure of the meson.

For the pseudoscalar mesons composed of the light $(u,d,s)$ quarks such as $(\pi^0,\eta,\eta')$, there have been many
experimental data for spacelike regions $(Q^2>0$) 
up to $Q^2\sim 40$ GeV$^2$~\cite{Behrend:1990sr,Gronberg:1997fj,Denig:2014mma,Uehara:2012ag,Aubert:2009mc,BABAR:2011ad}.
Especially, for the high $Q^2$, 
the TFFs can be calculated asymptotically at leading twist as a convolution of the perturbative 
hard scattering amplitude and the nonperturbative meson distribution 
amplitude (DA)~\cite{Lepage:1980fj,Efremov:1979qk,Chernyak:1983ej}.  One of the prominent features of the perturbative QCD (pQCD)
is that the TFFs show the asymptotic behaviors, 
 $Q^2F_{(\pi,\eta,\eta')\gamma}(Q^2)\to{\rm constant}$
 as  $Q^2\to\infty$.
However, the results $Q^2F_{\pi\gamma}(Q^2)$ from the BaBar Collaboration~\cite{Aubert:2009mc} are not only inconsistent with pQCD prediction but also show the rapid growth of $Q^2F_{\pi\gamma}(Q^2)$ for $Q^2>15$ GeV$^2$ while the measurement from
Belle Collaboration~\cite{Uehara:2012ag} are consistent with the asymptotic limit of QCD for $Q^2>15$ GeV$^2$.
On the other hand, the subsequent BaBar
data~\cite{BABAR:2011ad} for $Q^2F_{(\eta,\eta^{\prime})\gamma}(Q^2)$
provided a consistency with the pQCD prediction unlike the case of $Q^2F_{\pi\gamma}(Q^2)$.
These discrepancies for the results of $Q^2 F_{\pi\gamma}(Q^2)$  between the BaBar and
the Belle data  as well as for the different behaviors of the results between $Q^2 F_{\pi\gamma}(Q^2)$  and $Q^2F_{(\eta,\eta^{\prime})\gamma}(Q^2)$ for the high $Q^2$ region
have motivated many theoretical studies
~\cite{Mikhailov:2009kf,Radyushkin:2009zg,Polyakov:2009je,
Dorokhov:2013xpa,Wu:2010zc, Kroll:2010bf,Agaev:2012tm,
Roberts:2010rn,Brodsky:2011xx,Stefanis:2012yw,Lucha:2011if,
deMelo:2013zza,Agaev:2014wna}
to investigate the key issues for the resolution of discrepancies.

To examine the issue of the scaling behavior of $Q^2 F_{P\gamma}(Q^2)$ in the large $Q^2$, it may be necessary to analyze the corresponding form factor not only 
in the spacelike region but also in the timelike region. While there have been some theoretical analysis~\cite{Escribano:2015nra,Escribano:2015yup}
for the timelike region below the resonance value $q^2=m^2_P$
of meson $P$ with the physical mass $m_P$, we could not find any theoretical studies in timelike region for $q^2>m^2_P$.
The reason for the difficulty of analyzing the timelike region maybe
due to the singular nature and the complexity of the timelike form factor beyond the resonance region.
Nevertheless, in our recent work of the $(\pi^0,\eta,\eta')\to\gamma^*\gamma$ TFFs~\cite{Choi:2017zxn}, 
we have developed the new method to explore the timelike region without resorting to mere analytic continuation
from the spacelike region to the timelike region and analyzed the entire kinematic region (both for the timelike region and the spacelike region)
using the light-front quark model (LFQM)~\cite{Choi:1997iq,Choi:2007yu,Choi:1999nu,Choi:2007se,Choi:2009ai}.
Our direct calculation in timelike region shows the complete agreement not only with 
the analytic continuation result from the spacelike region but also with the result from the dispersion relation between
the real and imaginary parts of the form factor.
Our results of $Q^2 F_{(\pi,\eta,\eta')\gamma}(Q^2)$ were in good agreement with the available experimental data
for low $|Q^2|$ region and also consistent with the pQCD prediction for the high $|Q^2|$ region.

In this work, we explore the heavy quarkonia $(\eta_c, \eta_b)\to\gamma\gamma^*$ transitions in both spacelike and timelike regions
expanding our previous work of the $(\pi^0,\eta,\eta')\to\gamma^*\gamma$ TFFs~\cite{Choi:2017zxn}.
For the charmonium case,  the form factor $F_{\eta_c\gamma}(Q^2)$
was measured from BaBar collaboration~\cite{Lees:2010de} only in the spacelike region of 2 GeV$^2$ $<Q^2<$ 50 GeV$^2$.
There have been several theoretical studies on the TFF $F_{\eta_c\gamma}(Q^2)$ in the spacelike region
using various theoretical approaches and phenomenological models such as
pQCD~\cite{Feldmann:1997te,Cao:1997hw}, lattice QCD~\cite{Dudek:2006ut,Chen:2016yau}, 
non-relativistic QCD (NRQCD)~\cite{Feng-Jia-Sang,Wang-Wu-Sang-Brodsky}, QCD sum rules~\cite{Lucha:2012ar}, LFQM~\cite{Geng:2013yfa}, and covariant approach
based on Dyson-Schwinger and Bethe-Salpeter (BS) equations~\cite{Chen:2016bpj}.  
In particular, a strong discrepancy between the NRQCD prediction~\cite{Feng-Jia-Sang} and the BaBar measurements has been recently resolved by
applying the Principle of Maximum Conformality to the renormalization scale~\cite{Wang-Wu-Sang-Brodsky}.  
Also to overcome the weakness of the Dyson-Schwinger approach caused by a series of complex-valued singularities with increasing photon-momentum square in the numerical Euclidean momentum integration, a novel method using the perturbation theory integral representations of the quark propagator, meson amplitude and quark-photon vertex has been implemented to calculate the $F_{\eta_c\gamma}(Q^2)$ for any spacelike momenta~\cite{Chen:2016bpj}. In contrast to these and other available theoretical approaches and phenomenological 
models, the salient feature of our LFQM analysis is to explore the timelike region as well as the spacelike region within the same theoretical framework.
As we discuss in this work, the LFQM analysis of the TFF $F_{P\gamma} (Q^2)$ amenable 
both for the spacelike region ($Q^2 >0$) and the timelike region ($Q^2 <0$) provides a systematic twist expansion
of $Q^2 F_{P\gamma} (Q^2)$ for the high $|Q^2|$ region. 

The paper is organized as follows. In Sec.~\ref{sec:II}, we briefly discuss the TFFs  obtained from the $q^+(=q^0+q^3)\neq 0$ frame
in our LFQM starting from an exactly solvable covariant BS model of (3+1)-dimensional fermion field theory. 
The self-consistent correspondence relations  between the covariant BS model and our
LFQM are also discussed and the explicit form of $F_{(\eta_c, \eta_b)\gamma}(Q^2)$  in our LFQM  is presented. 
Especially, a systematic twist expansion of $Q^2 F_{(\eta_c, \eta_b)\gamma}(q^2)$ is provided explicitly and the leading- and higher-twist effects 
in the calculations of $Q^2 F_{(\eta_c, \eta_b)\gamma}(q^2)$ are discussed in this section.
In Sec.~\ref{sec:III}, we present our numerical results for the transverse momentum dependent distribution amplitude (TMDA), which is a 3-dimensional generalization of the DA, as well as its longitudinal and transverse moments. 
The $(\eta_c,\eta_b)\to\gamma^*\gamma$ TFFs for both spacelike and timelike regions are obtained 
and compared with the available experimental data. In order to check the validity of our LFQM calculations in the timelike regions, we verify
the exact agreement of our direct LFQM calculation in the timelike region with the results obtained from the dispersion relation between the
real and imaginary parts of the form factors.
Conclusions follow in Sec.~\ref{sec:V}. 

\section{Light-Front Quark Model Description}
%\subsection{Two-point function: decay amplitude}
\label{sec:II}
\begin{figure*}\label{fig1}
%\begin{center}
\includegraphics[height=3.5cm, width=13cm]{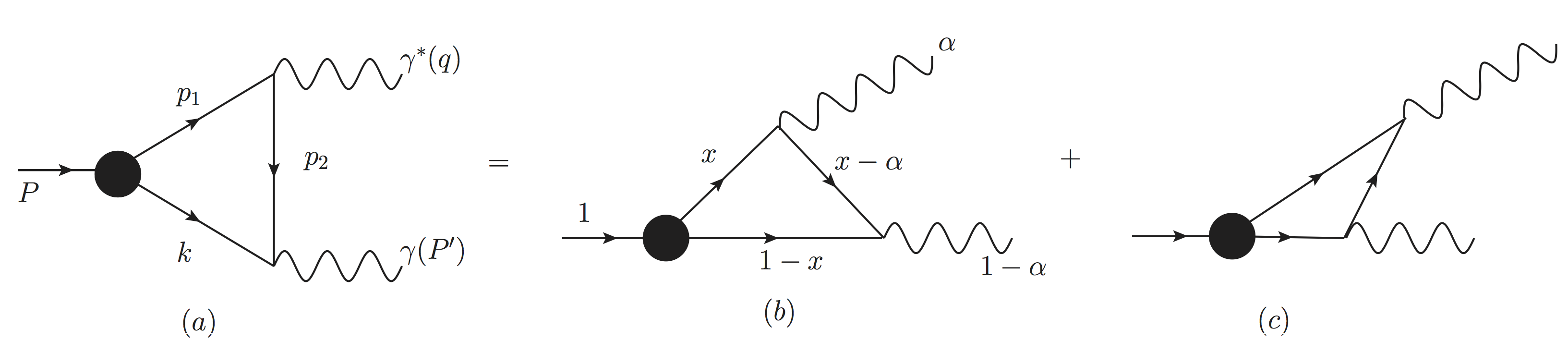}
\caption{
One-loop Feynman diagrams that contribute to $P\to\gamma^*\gamma$. The single covariant Feynman diagram (a) is in principle the same
as the sum of the two LF time-ordered diagrams (b) and (c), respectively. 
}
%\end{center}
\end{figure*}
The transition form factor $F_{P\gamma}$ for 
the $P\to\gamma^*\gamma$~($P=\pi^0, \eta, \eta',\eta_c,\eta_b$)
transition is defined from
the matrix element of electromagnetic current $\Gamma^\mu=\la\gamma(P-q)|J^\mu|P(P)\ra$
as follows:
\be\label{Eq1}
\Gamma^\mu = i e^2 F_{P\gamma}(Q^2)\ep^{\mu\nu\rho\sigma}P_\nu\vep_\rho q_\sigma,
\ee
where $P^\mu$ and $q^\mu$ are the four momenta of the incident pseudoscalar meson and virtual photon,
respectively, and $\vep$ is the transverse polarization vector of the final (on-shell)
photon. This process is illustrated by the Feynman diagram in Fig. 1 (a).
In the exactly solvable manifestly covariant BS model, 
the covariant amplitude $\Gamma^\mu$ is obtained by the following momentum integral
\be\label{Eq2}
\Gamma^\mu = i e_Q e_{\bar Q} N_c
\int\frac{d^4k}{(2\pi)^4} \frac{ {\rm Tr}\left[\gamma_5\left(\slash \!\!\!\!\!p_1 + m_Q \right)
 \gamma^\mu \left(\slash \!\!\!\!\!p_2 + m_Q \right)\slash \!\!\!\!\vep
 \left(-\slash \!\!\!\!k + m_Q \right) \right]}
{N_{p_1} N_k N_{p_2}}H_0,
\ee
where $N_c$ is the number of colors and $e_{Q(\bar Q)}$ is the quark~(antiquark) electric charge.
The denominators $N_{p_j} (= p_{j}^2 -m_Q^2 +i\ep) (j=1,2)$
and $N_k(= k^2 - m_{\bar Q}^2 + i\ep)$ come from the intermediate quark and antiquark propagators
of mass $m_Q=m_{\bar Q}$ carrying the internal
four-momenta $p_1=P-k$, $p_2=P-q-k$, and $k$, respectively.
The ${\bar q}q$ bound-state vertex function of the meson is denoted by $H_0$.

It is well known that the single covariant Feynman diagram Fig. 1 (a) is in general equal to the sum of the two LF time-ordered
diagrams Figs.~1 (b) and~1(c) if the $q^+\neq 0$ frame is taken. However, if the $q^+=0$ frame (but ${\bf q_\perp}\neq 0$ so 
that $q^2=q^+q^- -{\bf q}^2_\perp=-{\bf q}^2_\perp=-Q^2$) is chosen, the LF diagram 1(c) does not contribute but
only the diagram 1(b) gives exactly the same result as the covariant diagram 1(a). This has been known to be the virtue of
taking the $q^+=0$ frame in the LF calculation and many previous LF calculations have adopted this $q^+=0$ frame in the analysis 
of meson-photon TFFs~\cite{Lepage:1980fj,deMelo:2013zza,Choi:2007yu,Cao:1997hw}.
However, the analysis in the timelike region using the $q^+=0$ frame has been challenging 
since the $q^+=0$ frame is defined only in the spacelike region ($Q^2>0$) and the analytic continuation from 
spacelike region to timelike ($q^2=-Q^2>0$)  region is not quite straightforward due to the complication of mixture between 
the external momentum ${\bf q}_\perp$ and the internal
momentum ${\bf k}_\perp$  included in the term showing the singularity in the timelike region as discussed in~\cite{Choi:2017zxn}.

To overcome this difficulty in the analysis of the meson-photon TFFs in the timelike region, 
we recently explored in~\cite{Choi:2017zxn} the  $q^+\neq 0$ frames  (but with ${\bf q}_\perp=0$)
defined in the timelike region, i.e. $\alpha=q^+/P^+=1- P'^+/P^+$ frames with (1) $0<\alpha<1$  and (2) $\alpha=1$. 
For the $0<\alpha<1$ case, the covariant diagram in Fig.~1 (a) is shown to be equivalent to the sum of two LF diagrams 
Figs.~1 (b) and~1(c).
However, for the case of $\alpha=1$, we find that Fig. 1(b) does not contribute but only Fig. 1 (c) contributes 
to the total transition amplitude and coincides with the covariant result of Fig. 1(a).
The salient feature of the $\alpha=1$ frame not only show the boost invariant result but also show much more effective
computation of the timelike form factor over the commonly used  $q^+=0$ (i.e. $\alpha=0$) frame calculation~\cite{Choi:2017zxn}.
By applying the self-consistent correspondence relations (see, e.g., Eq. (35) in~\cite{Choi:2014ifm}) between the covariant BS model and our LFQM 
found in the analysis of the twist-2 and twist-3 DAs of
pseudoscalar and vector mesons~\cite{Choi:2014ifm,Choi:2013mda,Choi:2016meb} and the pion electromagnetic form factor~\cite{Choi:2014ifm},
we were able to obtain the meson-photon TFFs~\cite{Choi:2017zxn} using the $\alpha=1$ frame 
with the more phenomenologically accessible Gaussian wave functions backed by the 
LFQM analysis of meson mass spectra~\cite{Choi:1997iq,Choi:2007yu,Choi:1999nu,Choi:2007se,Choi:2009ai}.

Since the TFFs for the heavy quarkonina $(\eta_c, \eta_b)\to\gamma^*\gamma$ transitions have the same form as the $F_{\pi\gamma}$ 
in~\cite{Choi:2017zxn} apart from the charge factor,  we do not duplicate the same analysis here but display only the final form
of $F_{\eta_c(\eta_b)\gamma}$ obtained from the $\alpha=1$ frame in our LFQM:
\be\label{Eq3}
F_{\eta_c(\eta_b)\gamma}(q^2) = e^2_{c(b)} \frac{\sqrt{2 N_c}}{4\pi^3}\int^{1}_0
 \frac{dx}{(1-x)} \int d^2{\bf k}_\perp
 \frac{1}{M^2_0 - q^2}\Psi_{\frac{\uparrow\downarrow-\downarrow\uparrow}{\sqrt{2}}}(x,{\bf k}_\perp),
% \frac{\Phi_R(x,{\bf k}_{\perp})}{\sqrt{{\bf k}^2_\perp + m^2_Q}},
 \ee
where  $M^2_0 = \frac{ {\bf k}^{2}_\perp + m^2_Q}{x (1-x)}$ is the invariant mass and the LF wave function of a pseudoscalar meson with the constituent quark and antiquark mass $m_Q=m_{\bar Q}$ is given by
\be\label{Eq3-1}
\Psi_{\frac{\uparrow\downarrow-\downarrow\uparrow}{\sqrt{2}}}(x,{\bf k}_\perp)
=\frac{1}{\sqrt{2}}({\cal R}^{00}_{\uparrow\downarrow}-{\cal R}^{00}_{\downarrow\uparrow})\phi _{1S}(x,{\bf k}_\perp)
=  \frac{m_Q}{\sqrt{{\bf k}^2_\perp + m^2_Q}}\phi _{1S}(x,{\bf k}_\perp),
\ee
with the
spin-orbit wave function ${\cal R}^{JJ_z}_{\lambda_Q\lambda_{\bar{Q}}}$ obtained by the interaction independent Melosh transformation
from the ordinary equal-time static spin-orbit wave function assigned by the quantum number $J^{PC}$. Explicit form of 
${\cal R}^{00}_{\lambda_Q\lambda_{\bar{Q}}}$ for $m_Q=m_{\bar Q}$ case is given by
\be\label{Eq-2}
{\cal R}^{00}_{\lambda_Q\lambda_{\bar{Q}}}=\frac{1}{\sqrt{2}\sqrt{{\bf k}^2_\perp+m^2_Q}}
\begin{pmatrix}
-k^x + ik^y &  m_Q \\ -m_Q & -k^x - i k^y
\end{pmatrix},
\ee
which satisfies $\sum_{\lambda_Q\lambda_{\bar Q}} {\cal R}^{00\dagger}_{\lambda_Q\lambda_{\bar Q}}{\cal R}^{00}_{\lambda_Q\lambda_{\bar Q}}=1$.
For the radial wave function, we use in this work the  1$S$ state harmonic oscillator wave function
\be\label{QM2}
\phi_{1S}(x,{\bf k}_{\perp}) =
\frac{4\pi^{3/4}}{\beta^{3/2}} 
\sqrt{\frac{\partial k_z}{\partial x}} e^{-\frac{{\vec k}^2}{2\beta^2}},
%\\
%\phi_{2S}(x,{\bf k}_{\perp}) &=&
%\frac{4\pi^{3/4}}{\sqrt{6}\beta^{7/2}} (2{\vec k}^2 - 3\beta^2) 
%\sqrt{\frac{\partial k_z}{\partial x}} e^{-\frac{{\vec k}^2}{2\beta^2}},
\ee
where $\partial k_z/\partial x = M_0/4x(1-x)$  is the Jacobian of the variable transformation
$\{x,{\bf k}_\perp\}\to {\vec k}=({\bf k}_\perp, k_z)$
and $\beta$ is the variational parameter
fixed by our previous analysis of meson mass spectra~\cite{Choi:1997iq,Choi:1999nu,Choi:2009ai,Choi:2007se}.
In particular, ${\vec k}^2$ is given by ${\vec k}^2 = {\bf k}^2_\perp +k^2_z$ where $k_z = (x-1/2)M_0$.
The normalization of $\phi_{1S}$ is thus given by
\be\label{QM6_norm}
\int^1_0 dx \int\frac{d^2{\bf k}_\perp}{16\pi^3}
|\phi_{1S}(x,{\bf k}_{\perp})|^2=1.
\ee

We should note that the TFF in the $q^+=0$ frame is obtained by the following 
replacement of the denominator factor, $(M^2_0 - q^2)^{-1}\to [M'^2_{0}]^{-1}$ in Eq.~(\ref{Eq3}), where 
$M'_0 = M_0 ({\bf k}_\perp\to {\bf k}_\perp + (1-x){\bf q}_\perp)$ (see ~\cite{Choi:2017zxn} for more detailed derivation).
Compared to the pole structure $[M'^2_0]^{-1}$ in the timelike region of the $q^+=0$ frame,
the internal transverse momentum ${\bf k}_\perp$ for the corresponding pole structure $(M^2_0 - q^2)^{-1}$ in the $\alpha=1$ frame 
as shown in Eq.~(\ref{Eq3})
does not mix with the external virtual photon momentum $q$. Because of this salient feature for the $\alpha=1$ frame, the direct timelike
TFF calculation can be done most effectively in contrast to the computation in the $q^+=0$ frame. 
We have already explicitly shown in our numerical calculations~\cite{Choi:2017zxn}  for the  $(\pi^0,\eta,\eta')\to\gamma^*\gamma$ TFFs
that our direct results 
of the timelike form factors given by Eq.~(\ref{Eq3})  satisfy the following dispersion
relations (DR);
%between ${\rm Re}[F(q^2)]$ and ${\rm Im} [F(q^2)]$ 
%given by
\bea\label{ReF}
{\rm Re}\; F(q^2)&=&\frac{1}{\pi} P\int^\infty_{-\infty}\frac{ {\rm Im}\;F(q'^2)}{q'^2 -q^2} dq'^2,
\nonumber\\
{\rm Im}\; F(q^2)&=& -\frac{1}{\pi} P\int^\infty_{-\infty}\frac{ {\rm Re}\;F(q'^2)}{q'^2 -q^2} dq'^2,
\eea
where $P$ indicates the Cauchy principal value. 

Moreover, a systematic twist expansion of
$F_{\eta_c(\eta_b)\gamma}(q^2)$ is straightforwardly attained as discussed below by expanding the factor $1/(M_0^2 - q^2)$ in
geometric sum for high $Q^2=-q^2$;
\be\label{geometric_sum}
\frac{1}{M_0^2 - q^2}  = \frac{1}{M^2_0+Q^2}=\frac{1}{Q^2 (1 + \frac{M^2_0}{Q^2})}= \frac{1}{Q^2} -\frac{M^2_0}{Q^4}+\cdots.
\ee
With the expansion of the geometric sum given by Eq.~(\ref{geometric_sum}), 
we can easily expand $Q^2 F_{\eta_{c(b)}\gamma}(Q^2)$  in Eq.~(\ref{Eq3}) in
terms of  the twist-2, twist-3 DAs, etc. as follows
\be\label{Q2F}
Q^2 F_{\eta_c(\eta_b)\gamma}(q^2) = e^2_{c(b)}  f_M\int^1_0\frac{dx}{1-x} 
\biggl[ 2 \phi_{2;M} (x)  - 4 \frac{m_Q}{Q^2}\mu_M \phi_{3;M} (x)  + {\cal O}\biggl(\frac{1}{Q^{2n}}\biggr)
\biggr],
\ee
with $n\geq 2$. 
The normalized twist-2 DA $\phi_{2;M}(x)$ and twist-3 DA $\phi_{3;M}(x)$ for the meson $M(=\eta_c,\eta_b)$ obtained from
our LFQM
are given by~\cite{Choi:2014ifm}
\be\label{2DA}
\phi_{2;M} (x) = \frac {\sqrt{2N_c}} {f_M 8\pi^3} \int d^2{\bf k}_\perp
\Psi_{\frac{\uparrow\downarrow-\downarrow\uparrow}{\sqrt{2}}}(x,{\bf k}_\perp),
\ee
and
\be\label{3DA}
\phi_{3;M} (x) = \frac {\sqrt{2N_c}} {f_M \mu_M 16\pi^3} \int d^2{\bf k}_\perp
\biggl(\frac{M^2_0}{m_Q}\biggr)
\Psi_{\frac{\uparrow\downarrow-\downarrow\uparrow}{\sqrt{2}}}(x,{\bf k}_\perp),
%\frac{\Phi_{R}(x, {\bf k}_\perp)}{\sqrt{ {\bf k}^2_\perp + m_Q^2} }{M^2_0},
\ee
where $f_M$ is the decay constant and the normalization parameter $\mu_M$ in Eq.~(\ref{3DA})
results from quark condensate and can be fixed from the normalization of the DAs via
$\int^1_0 dx \;\phi_{2(3);M}(x) =1$. We should note that the twist-2 and twist-3 DAs $\phi_{2;M}$ and $\phi_{3;M}$ correspond
to the axial-vector and pseudoscalar channels of a meson M, respectively, as discussed in~\cite{Choi:2014ifm}.
The TFF for $\pi^0\to\gamma\gamma^*$  can be obtained by replacing the charge factor $e^2_{c(b)}$ in Eq.~(\ref{Q2F})
with $(e^2_u - e^2_d)/\sqrt{2}$. 
The form factor at zero momentum transfer is related with the decay width for $P\to\gamma\gamma$ via
\be\label{DW1}
\Gamma_{P\to\gamma\gamma} =\frac{\pi}{4}\alpha^2M_P^3|F_{P\gamma}(0)|^2,
\ee
where $\alpha$ is the fine structure constant and $M_P$ is the physical meson mass.

\section{Numerical Results}
\label{sec:III}

\begin{table}[t]
\begin{center}
\caption{Model parameters $(m_Q,\beta_{Q{\bar Q}})(Q=c,b)$ (in GeV).}
\label{t1}
\begin{tabular}{lcccccc} \hline\hline
Model & $m_c$ & $m_b$ & $\beta_{c{\bar c}}$ & $\beta_{b{\bar b}}$ & $f_{\eta_c}$ & $f_{\eta_b}$ \\
\hline
Set I & 1.80~ & 5.20~ & 0.6509~ & 1.1452~ & 0.326 & 0.507 \\
Set II  & 1.30 &  4.50 & 0.6509 & 1.1452 & 0.335 & 0.530  \\
Exp.~\cite{Ed} & - &  - & - & - & $0.335(75)$ & -  \\
\hline\hline
\end{tabular}
\end{center}
\end{table}
In our numerical calculations, we use the two sets of model parameters
for $\eta_c$ and  $\eta_b$ as shown in Table~\ref{t1}.  While the Set I
was obtained from the variational principle for the QCD-motivated effective Hamiltonian including the
linear confining potential and the hyperfine interaction~\cite{Choi:1997iq,Choi:2007yu,Choi:1999nu,Choi:2007se,Choi:2009ai}, the Set II
provides the parameter sensitivity check of our LFQM to the constituent quark masses and at the same time
the better fit to the experimental data for $f_{\eta_c}$~\cite{Ed} and $\Gamma_{\eta_c\to\gamma\gamma}$~\cite{PDG2018} . 
We should note that the TFFs are much more sensitive to the variation of the quark masses than to the variation
of the $\beta$ parameters.

Defining the transverse momentum dependent DA (TMDA)
 $\psi_{2(3);M}(x,{\bf k}_\perp)$ that is a 3-dimensional 
generalization of the twist-2(3) DA $\phi_{2(3);M}(x)$ as
\be\label{Eq7}
\phi_{2(3);M}(x) =\int^\infty_0 d^2{\bf k}_\perp\; \psi_{2(3);M}(x,{\bf k}_\perp)=\int^1_0 dy \; \psi_{2(3);M}(x,y),
\ee
the $n$th transverse moment is obtained by
\be\label{Eq8}
\la {\bf k}^n_{\perp}\ra_{2(3);M}=\int^1_0 dx \int^\infty_0 d^2{\bf k}_\perp\; \psi_{2(3);M}(x,{\bf k}_\perp) {\bf k}^n_\perp,
\ee
where $\psi_{2(3);M}(x,y)$ in Eq.~ (\ref{Eq7}) is obtained by changing the
variable ${\bf k}^2_\perp=y/(1-y)$. One can also define the expectation value of the longitudinal momentum,
so-called $\xi(=2x-1)$-moments, as follows:
\be\label{Eq9}
\la \xi^n_{\perp}\ra_{2(3);M}=\int^1_0 dx; \phi_{2(3);M}(x) \xi^n.
\ee

Our results of the 2nd transverse moment corresponding to the $\psi_{2;M}(x,{\bf k}_\perp)$
and $\psi_{3;M}(x,{\bf k}_\perp)$ wave functions
obtained from the Set I [Set II] are
$\la {\bf k}^2_{\perp}\ra_{2;\eta_c}=(866\;{\rm MeV})^2$ $[(840\;{\rm MeV})^2]$ and
$\la {\bf k}^2_{\perp}\ra_{3;\eta_c}=(940\;{\rm MeV})^2$ $[(950\;{\rm MeV})^2]$
for the $\eta_c$ meson
and $\la {\bf k}^2_{\perp}\ra_{2;\eta_b}=(1.573\;{\rm GeV})^2$ $[(1.561\;{\rm GeV})^2]$ and
$\la {\bf k}^2_{\perp}\ra_{3;\eta_b}=(1.636\;{\rm GeV})^2$ $[(1.640\;{\rm GeV})^2]$ 
for the $\eta_b$ meson, respectively. The 2nd $\xi$-moments of the twist-2 and twist-3 DAs
obtained from the Set I [Set II] are
$\la \xi^2 \ra_{2;\eta_c}= 0.0766~[0.111]$ and 
$\la \xi^2 \ra_{3;\eta_c}= 0.0859~[0.128]$ for the $\eta_c$ meson
and
$\la \xi^2 \ra_{2;\eta_b}= 0.0377~[0.0471]$ 
and
$\la \xi^2 \ra_{3;\eta_b}= 0.0402~[0.0510]$
for the $\eta_b$ meson, respectively.
\begin{figure}\label{fig2}
\begin{center}
\includegraphics[width=0.32\columnwidth,clip=]{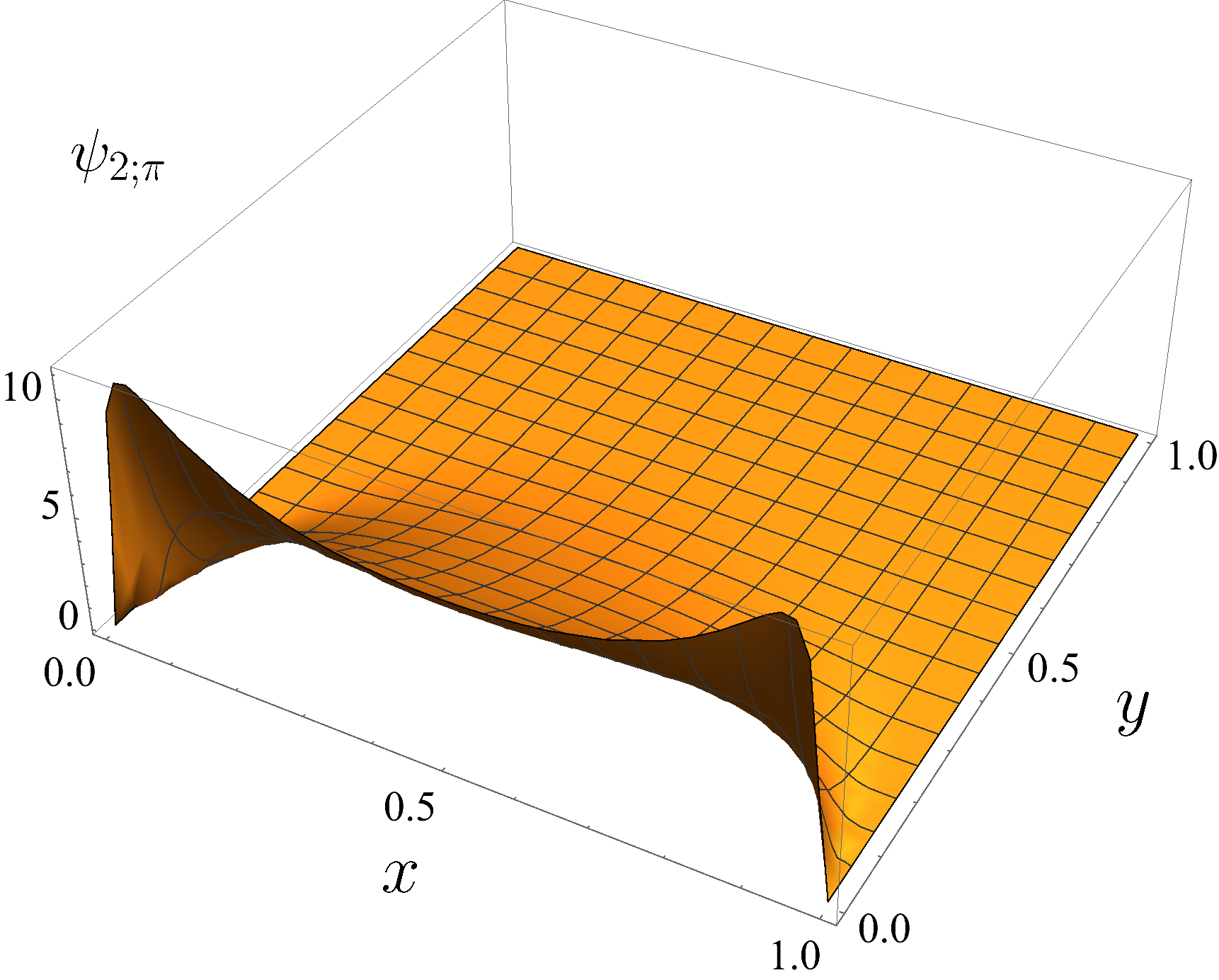}
\includegraphics[width=0.32\columnwidth,clip=]{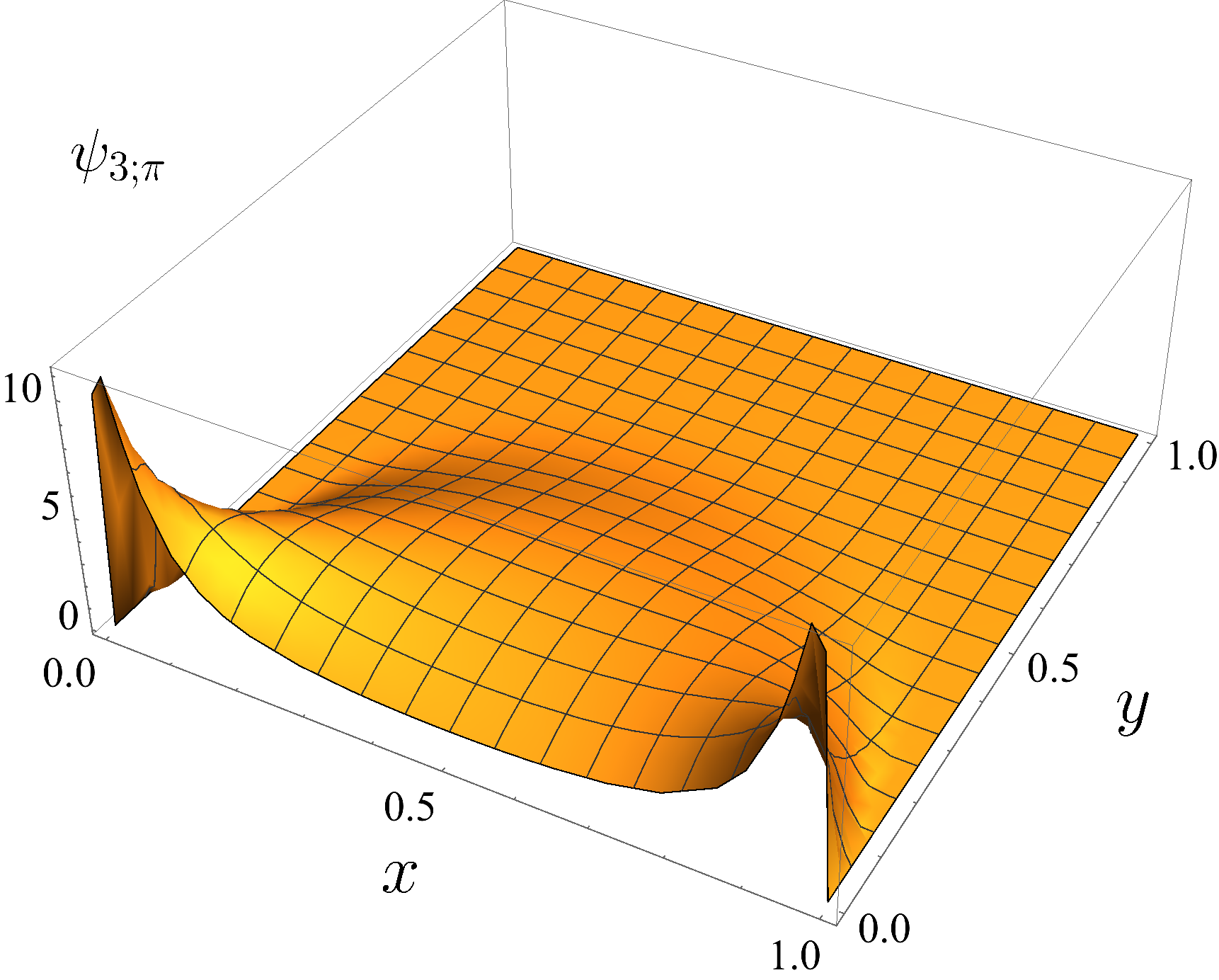}
\\
\includegraphics[width=0.32\columnwidth,clip=]{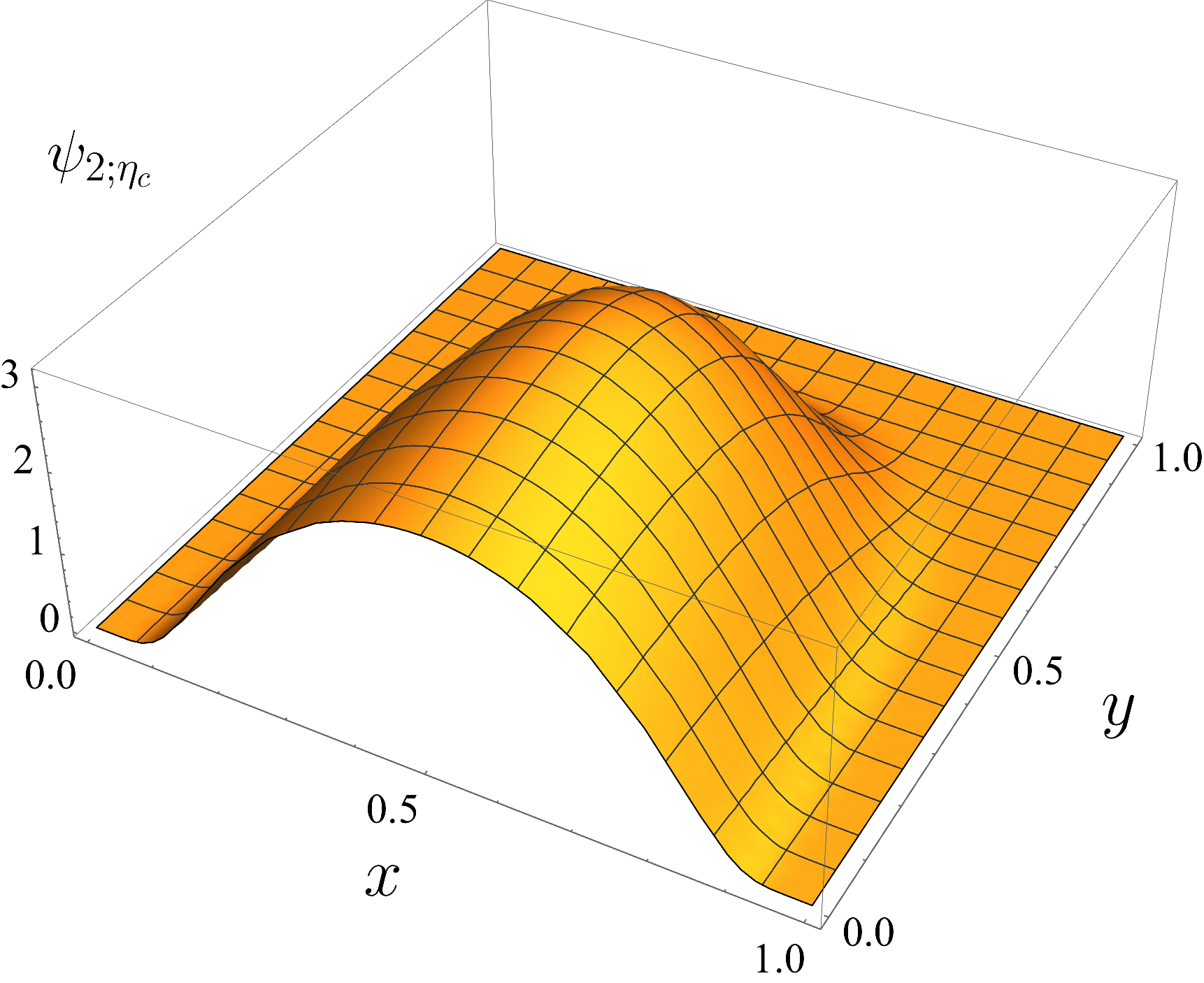}
\includegraphics[width=0.32\columnwidth,clip=]{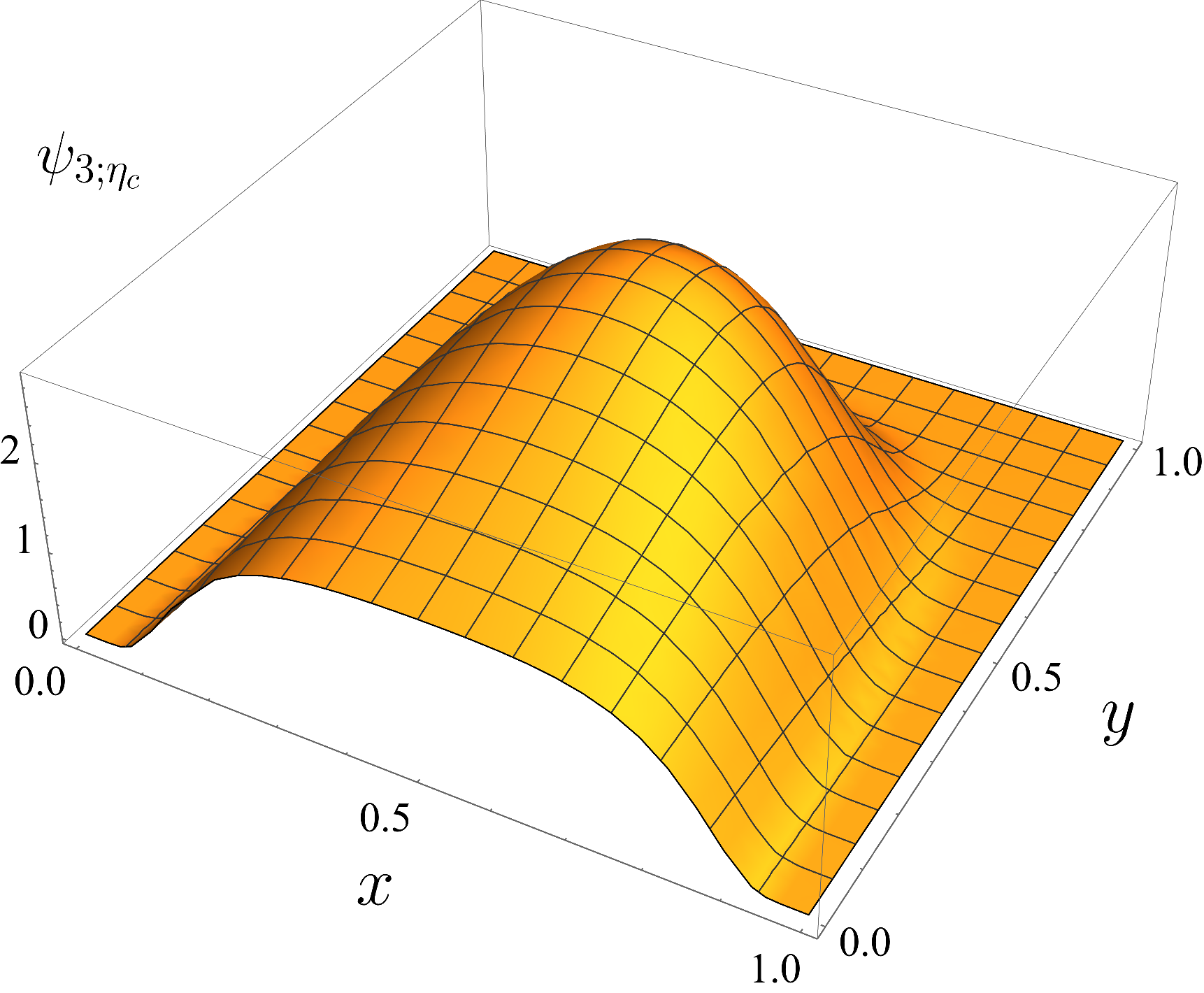}
\\
\includegraphics[width=0.32\columnwidth,clip=]{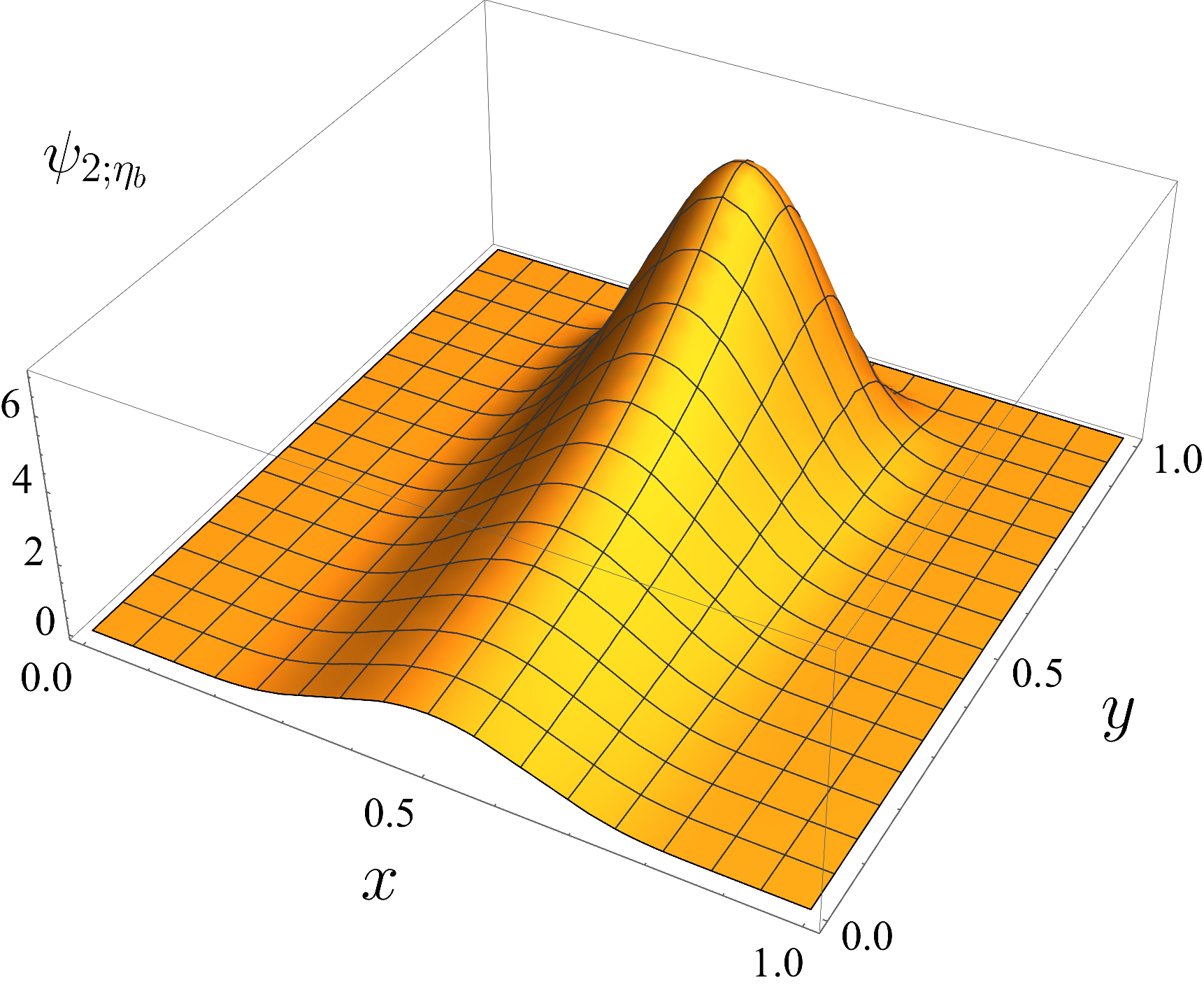}
\includegraphics[width=0.32\columnwidth,clip=]{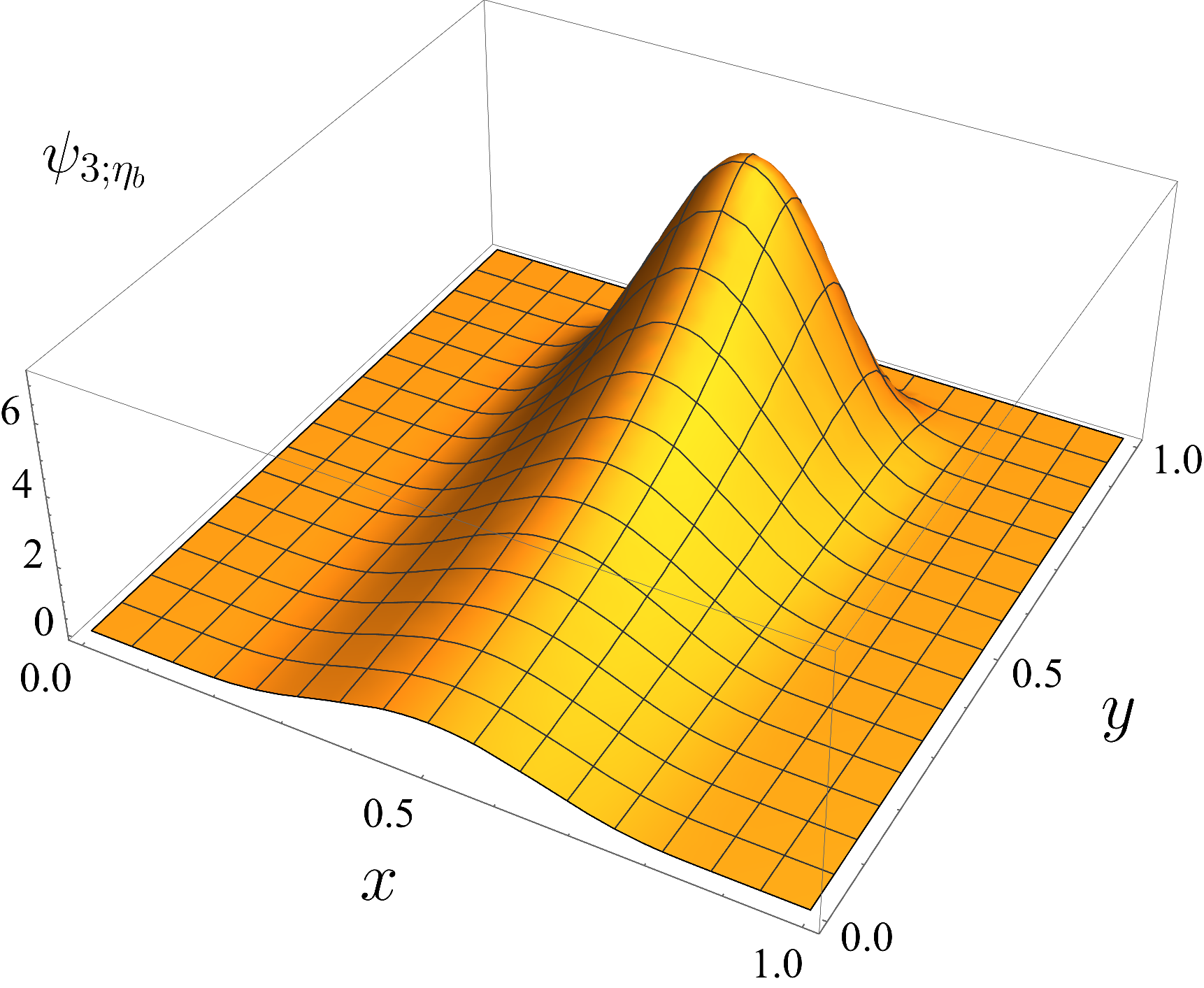}
\caption{Transverse momentum dependent distribution amplitudes 
(TMDAs) $\psi_{2;\pi}(x,y)$ (left panel)
 and $\psi_{3;\pi}(x,y)$ (right panel) for the $\pi$ meson (upper panel), 
$\psi_{2;\eta_{c}}(x,y)$ (left panel)
 and $\psi_{3;\eta_{c}}(x,y)$ (right panel) for the $\eta_c$ meson (middle panel),
 and $\psi_{2;\eta_{b}}(x,y)$ (left panel) 
 and $\psi_{3;\eta_{b}}(x,y)$ (right panel) for the $\eta_b$ meson (lower panel) 
obtained from the Set II, respectively.}
\end{center}
\end{figure}
%
%The twist-2 DAs $\phi_{\eta_{c(b)}}(x)$  obtained from the Set I can be found in~\cite{Choi:2009ai}.

Figure~2 shows the TMDAs of $\eta_{c(b)}$ related with the twist-2 and 3 DAs compared with those of $\pi$ meson obtained in our
previous work~\cite{Choi:2014ifm}, i.e.,
$\psi_{2;\pi}(x,y)$ (left panel)
 and $\psi_{3;\pi}(x,y)$ (right panel) for the $\pi$ meson (upper panel), 
$\psi_{2;\eta_{c}}(x,y)$ (left panel)
 and $\psi_{3;\eta_{c}}(x,y)$ (right panel) for the $\eta_c$ meson (middle panel),
 and $\psi_{2;\eta_{b}}(x,y)$ (left panel) 
 and $\psi_{3;\eta_{b}}(x,y)$ (right panel) for the $\eta_b$ meson (lower panel) 
obtained from the Set II, respectively.
Comparing the TMDAs $\psi_{2;M}(x,y)$ related with the twist-2 DAs and $\psi_{3;M}(x,y)$ related with the twist-3 DAs, we find that
$\psi_{3;M}(x,y)$ shows in general broader shape and receives higher ${\bf k}_\perp$-contributions than $\psi_{2;M}(x,y)$ regardless
of the kinds of mesons $M(=\pi, \eta_c,\eta_b)$. 
We note
the reason why $\eta_b$ twist-2 and 3-contributions look so similar is due to such a large $b$-quark mass. 
On the other hand, as one can see from Fig.~2,  $\psi_{2(3);\pi}(x,y)$ receives contributions from the end points of $x$ for small ${\bf k}_\perp$ regions 
more than the heavy quarkonia case. We also note that 
$\psi_{2(3);\eta_{b}}(x,y)$ not only show much narrower shapes 
but also receive higher ${\bf k}_\perp$-contributions than $\psi_{2(3);\eta_{c}}(x,y)$ and  $\psi_{2(3);\pi}(x,y)$.
For the case of heavy quarkonia TMDAs, the results from the Set I are qualitatively very similar to those from the Set II
but show slightly narrower shape than those from the Set II due to the heavier quark masses.
As was discussed in~\cite{Choi:2017zxn},  we can associate the scale $\mu$, which separates nonperturbative and perturbative regimes,
with the transverse integration cutoff via $|{\bf k}_\perp|\leq \mu$. Since the twist-2 and twist-3 TMDAs  for heavy quarkonia show the higher ${\bf k}_\perp$
contributions than those for the pion, one can easily see the scale gets larger for the heavier quark. For the case of twist-2 TMDAs shown in 
Fig.~2,  we find that the integrations up to $y\simeq (0.5, 0.8, 0.93)$ of $\psi_{2; (\pi,\eta_c,\eta_b)}(x,y)$ make up 99$\%$ of the full results
for $\phi_{2, (\pi,\eta_c,\eta_b)}(x)$, respectively. This implies that our cutoff scales correspond to $\mu\simeq|{\bf k}_\perp|\simeq (1, 2,  3.6)$ GeV for the calculations
of the twist-2 $\phi_{2,(\pi,\eta_c,\eta_b)}(x)$, respectively.

The TFFs at $Q^2=0$ are obtained as $F_{\eta_c\gamma}(0)=0.0374~ [0.0664]$ GeV$^{-1}$ and $F_{\eta_b\gamma}(0)=0.0019 ~[0.0026]$ GeV$^{-1}$
for the Set I [Set II], respectively. Using the following experimental values of $(M_{\eta_c}, M_{\eta_b})=(2.98, 9.40)$ GeV~\cite{PDG2018}, we obtain
$\Gamma_{\eta_c\to\gamma\gamma}=1.55~ [4.88]$ keV and
$\Gamma_{\eta_b\to\gamma\gamma}=0.128 ~[0.239]$ keV 
for the Set I [Set II], respectively. The experimental value of $F_{\eta_c\gamma}(0)$ may be obtained from the experimental data
$\Gamma^{\rm exp}_{\eta_c\gamma\gamma}=5.1\pm 0.4$ keV~\cite{PDG2018}, which yields $F^{\rm exp}_{\eta_c\gamma\gamma}=0.067\pm0.0028$ GeV$^{-1}$.
Although our LFQM result for  $F_{\eta_c\gamma}(0)$ obtained from the Set II rather than the Set I shows a good agreement with the experimental value, 
 we should note that a recent lattice QCD result~\cite{Chen:2016yau} of 
 %$\Gamma_{\eta_c\to\gamma\gamma}=1.122(14)$ keV 
$F_{\eta_c\gamma}(0)=0.0318(2)$
corresponding to $\Gamma_{\eta_c\to\gamma\gamma}=1.122(14)$ keV
is similar to ours obtained from the Set I. 
\begin{figure*}\label{fig3}
\begin{center}
\includegraphics[width=0.35\columnwidth,clip=]{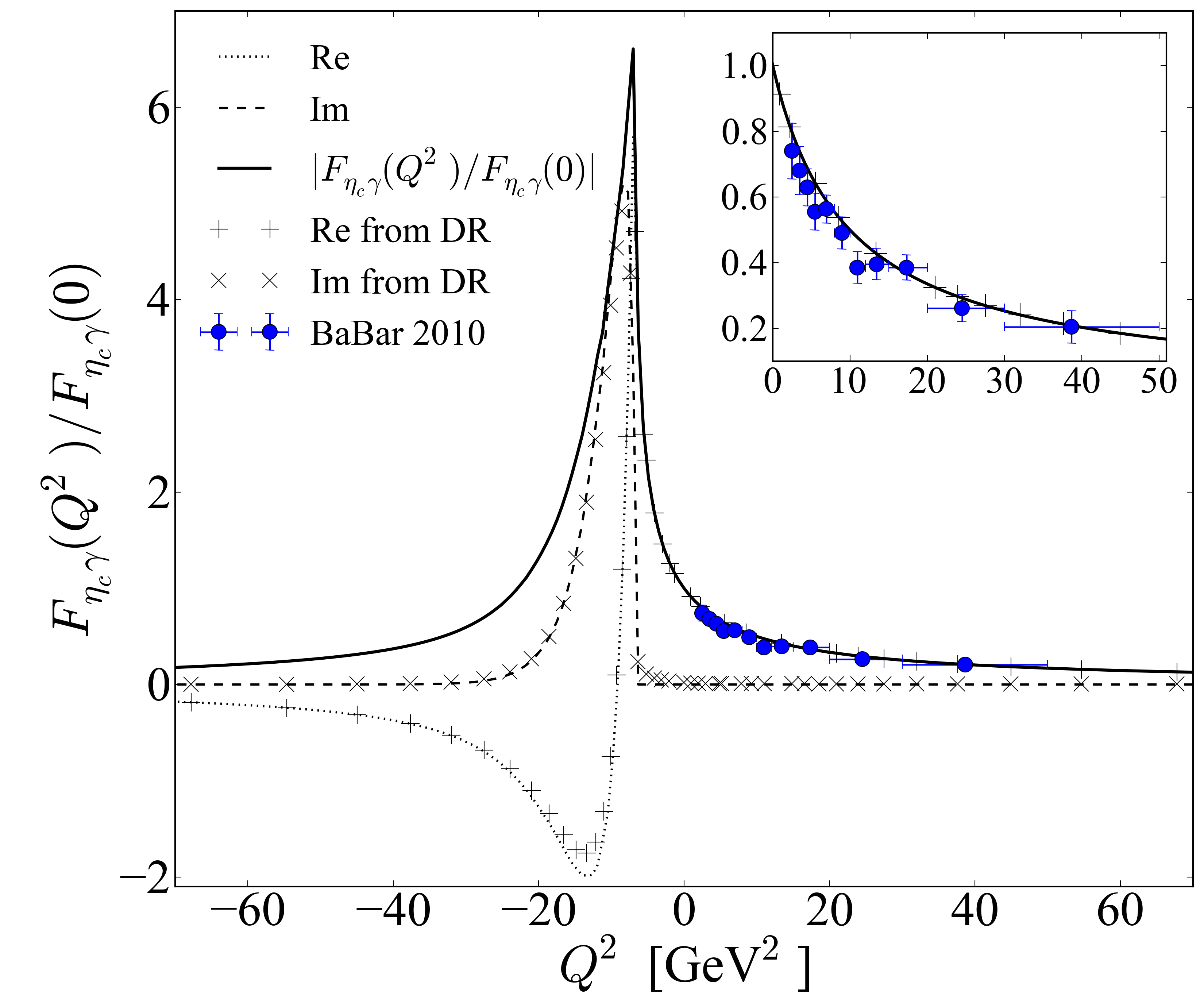}
\caption{The normalized $\eta_c\to\gamma\gamma^*$ transition form factor $F_{\eta_c\gamma}(Q^2)/F_{\eta_c\gamma}(0)$ obtained from the Set II
for both timelike ($q^2=-Q^2>0$)
spacelike ($q^2=-Q^2<0$) momentum transfer regions compared with the results obtained from the dispersion relation (DR). The data are taken from~\cite{Lees:2010de}.
}
\end{center}
\end{figure*}
\begin{figure*}\label{fig4}
%\begin{center}
\includegraphics[width=0.35\columnwidth,clip=]{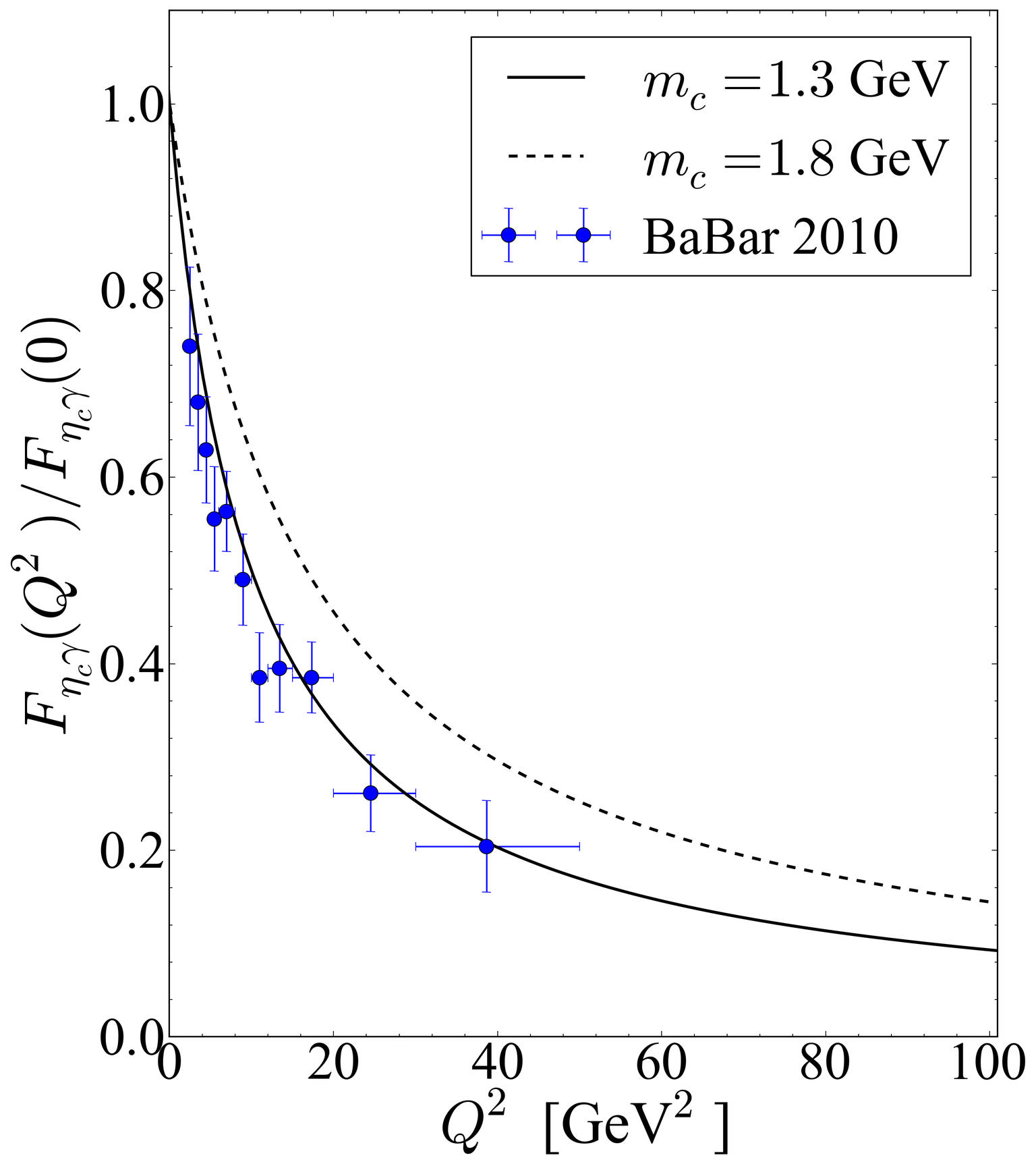}
%\hspace{1cm}
\includegraphics[width=0.37\columnwidth,clip=]{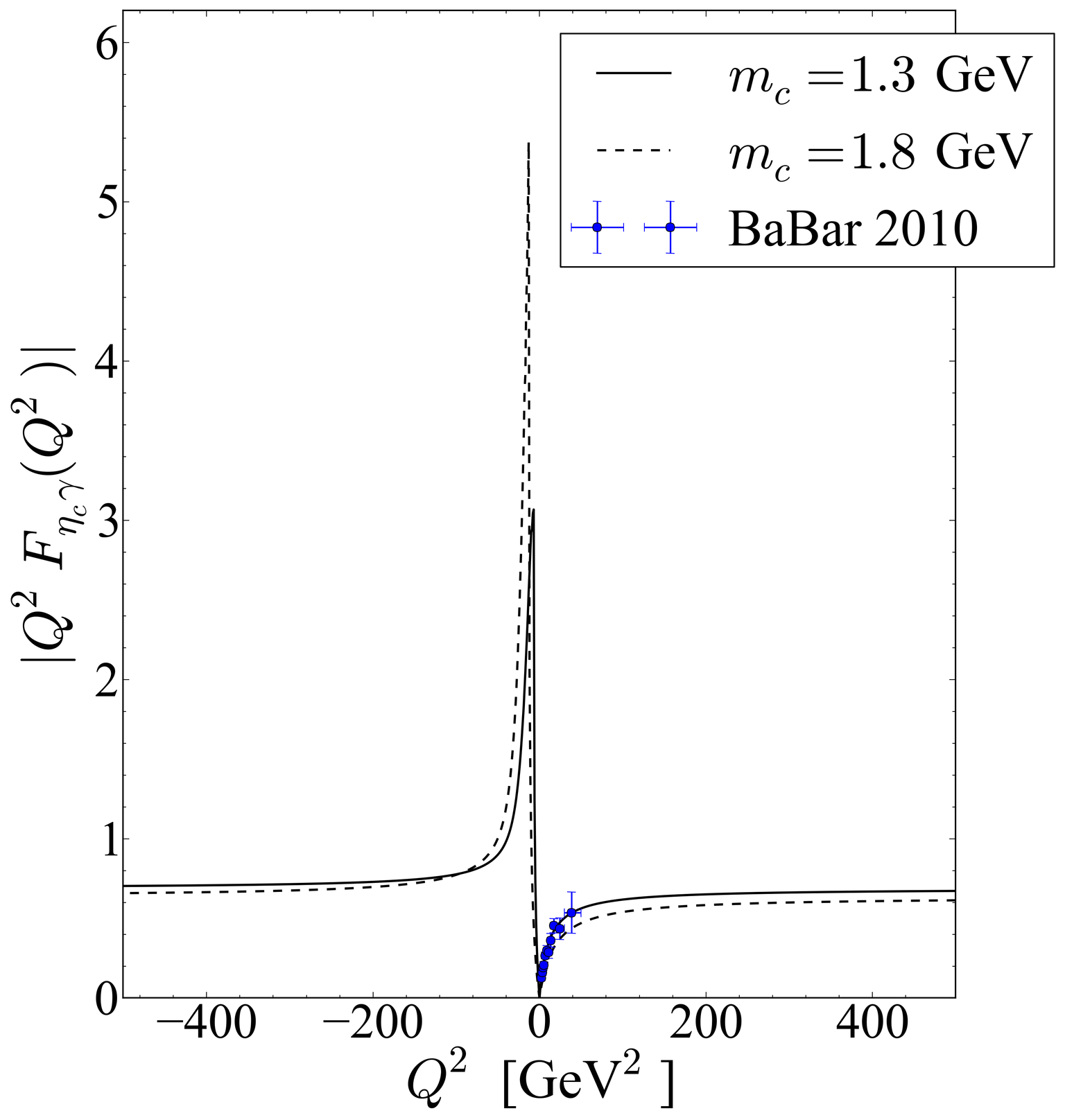}
\caption{The normalized $\eta_c\to\gamma\gamma^*$ transition form factor $F_{\eta_c\gamma}(Q^2)/F_{\eta_c\gamma}(0)$ in the
spacelike ($q^2=-Q^2<0$) momentum transfer region (left panel), and  the $|Q^2F_{\eta_c\gamma}(Q^2)|$ for both timelike ($q^2>0$) and spacelike 
momentum transfer regions (right panel). The data are taken from~\cite{Lees:2010de}.
}
%\end{center}
\end{figure*}
\begin{figure*}\label{fig5}
%\begin{center}
\includegraphics[width=0.35\columnwidth,clip=]{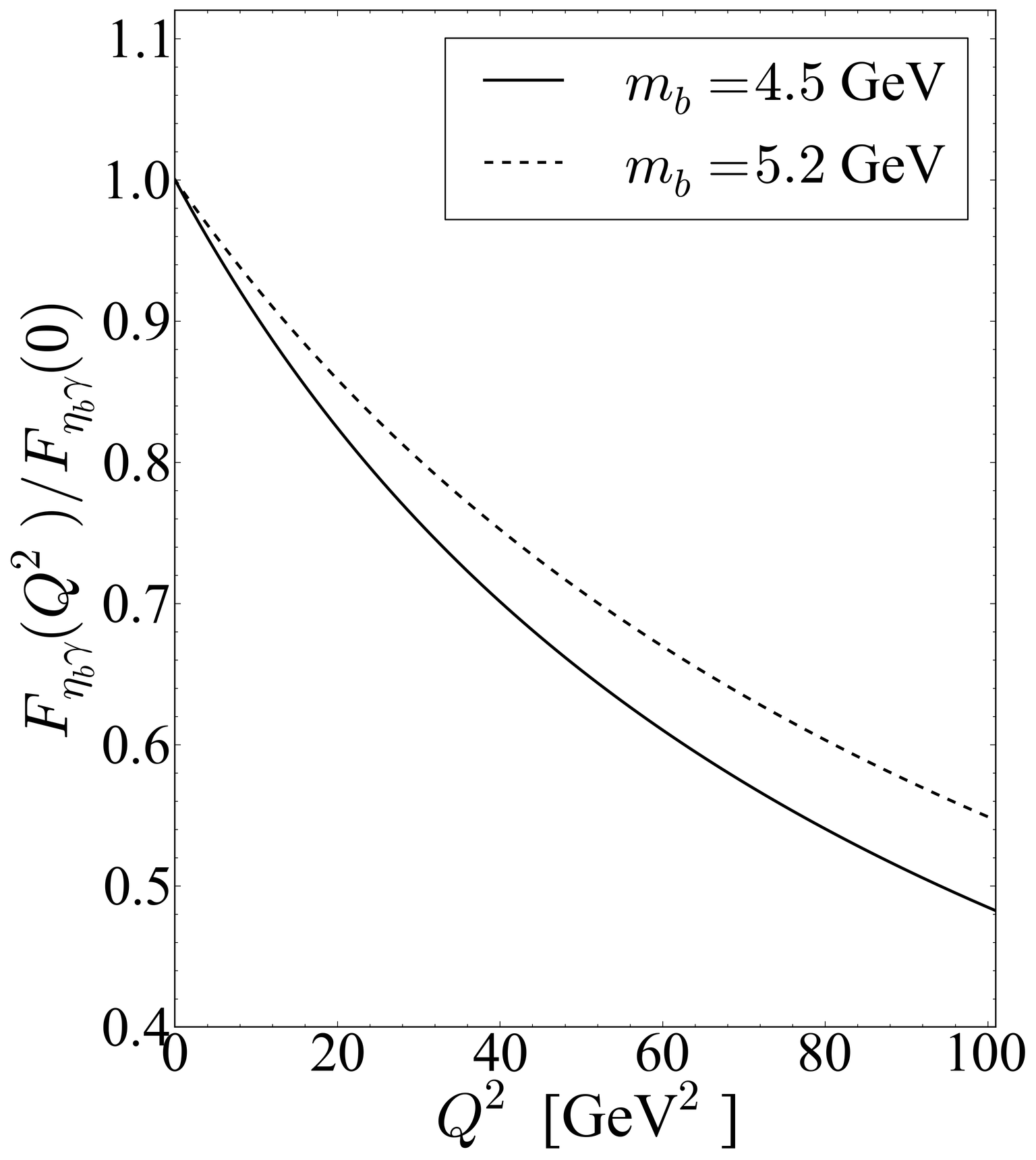}
%\hspace{1cm}
\includegraphics[width=0.37\columnwidth,clip=]{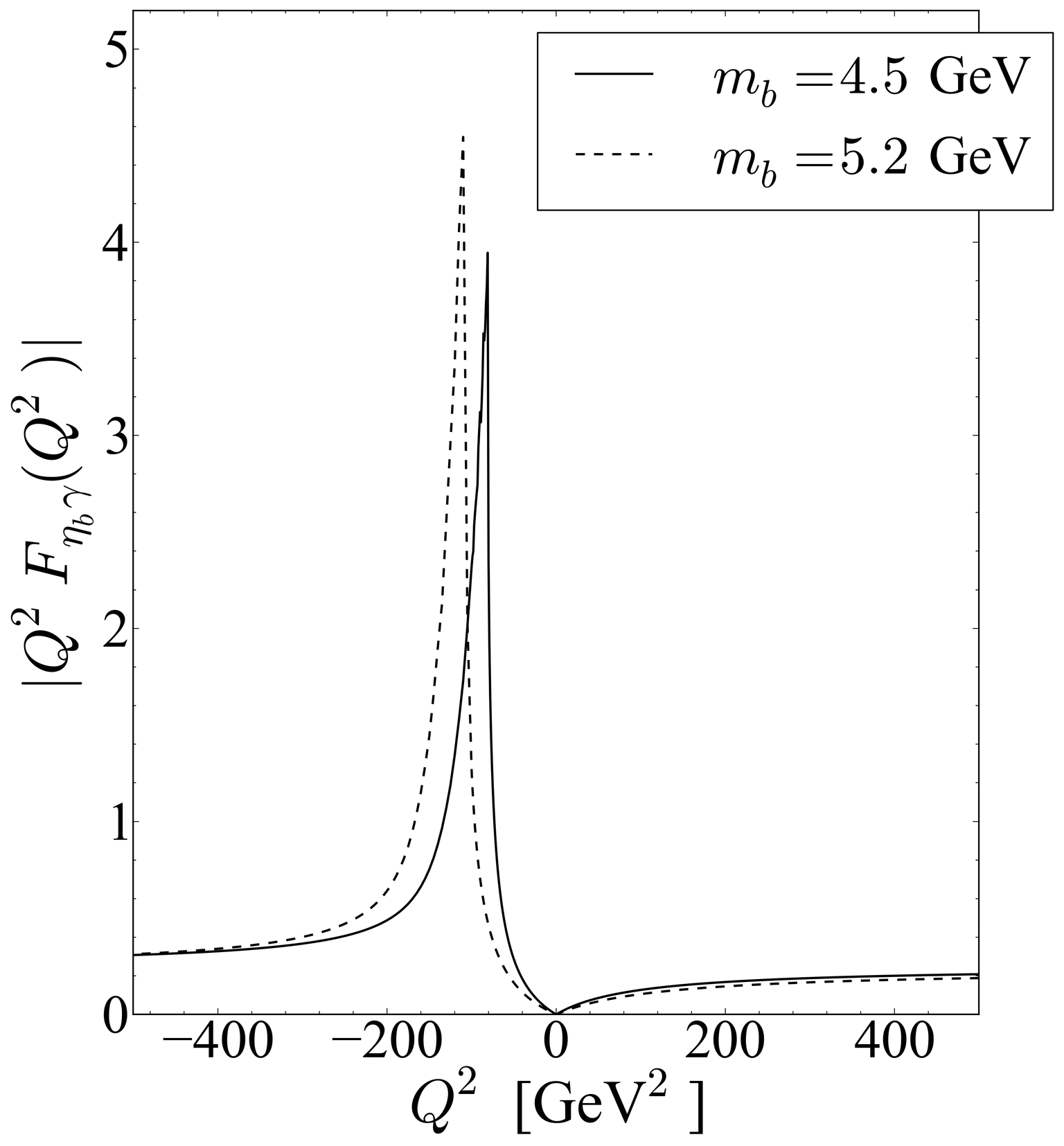}
\caption{The normalized $\eta_b\to\gamma\gamma^*$ transition form factor $F_{\eta_b\gamma}(Q^2)/F_{\eta_b\gamma}(0)$ in the
spacelike  momentum transfer region (left panel), and the $|Q^2F_{\eta_b\gamma}(Q^2)|$ for both timelike  and spacelike 
momentum transfer regions (right panel). 
}
%\end{center}
\end{figure*}
\begin{figure*}\label{fig6}
%\begin{center}
\includegraphics[width=0.37\columnwidth,clip=]{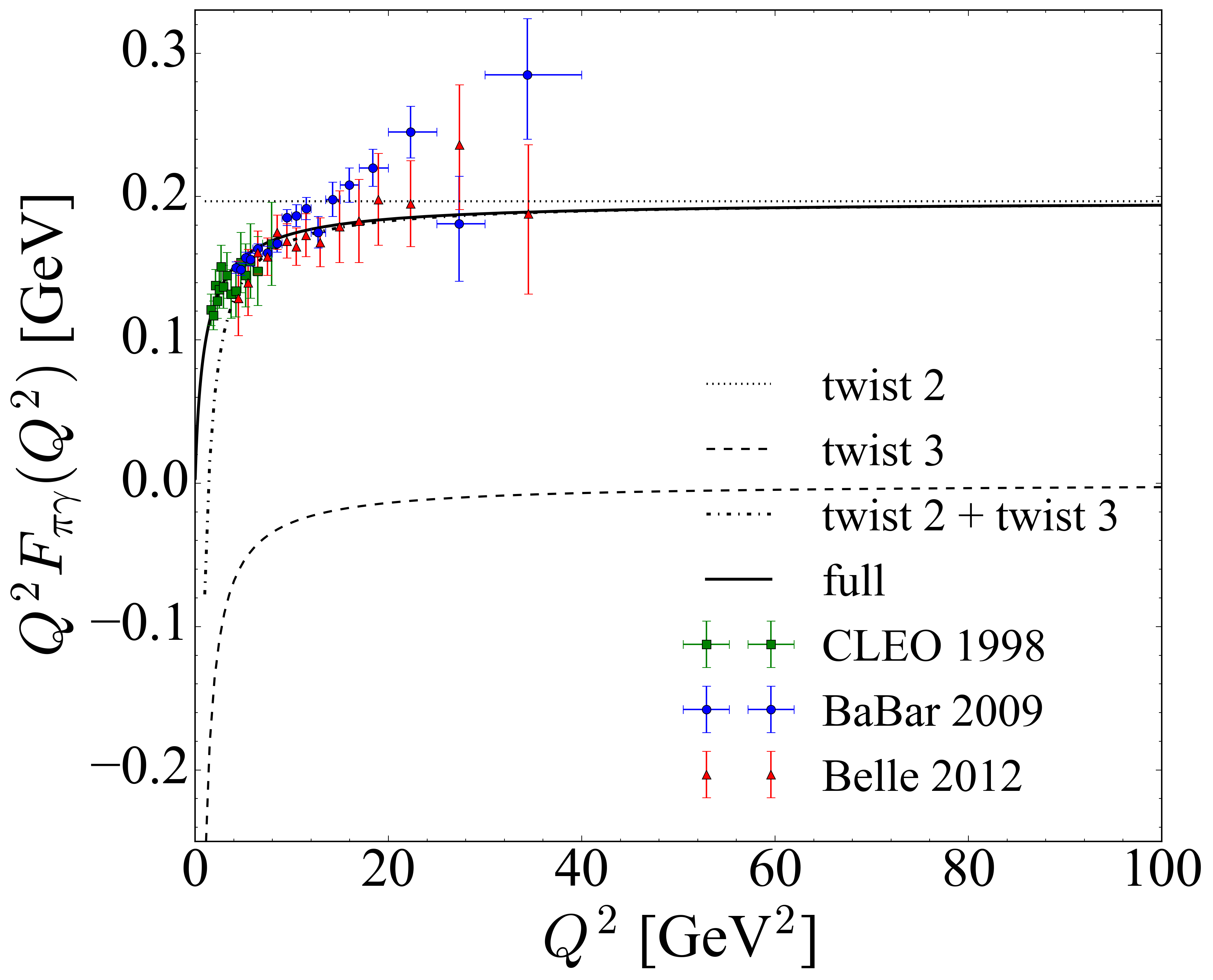}
%\hspace{1cm}
\includegraphics[width=0.37\columnwidth,clip=]{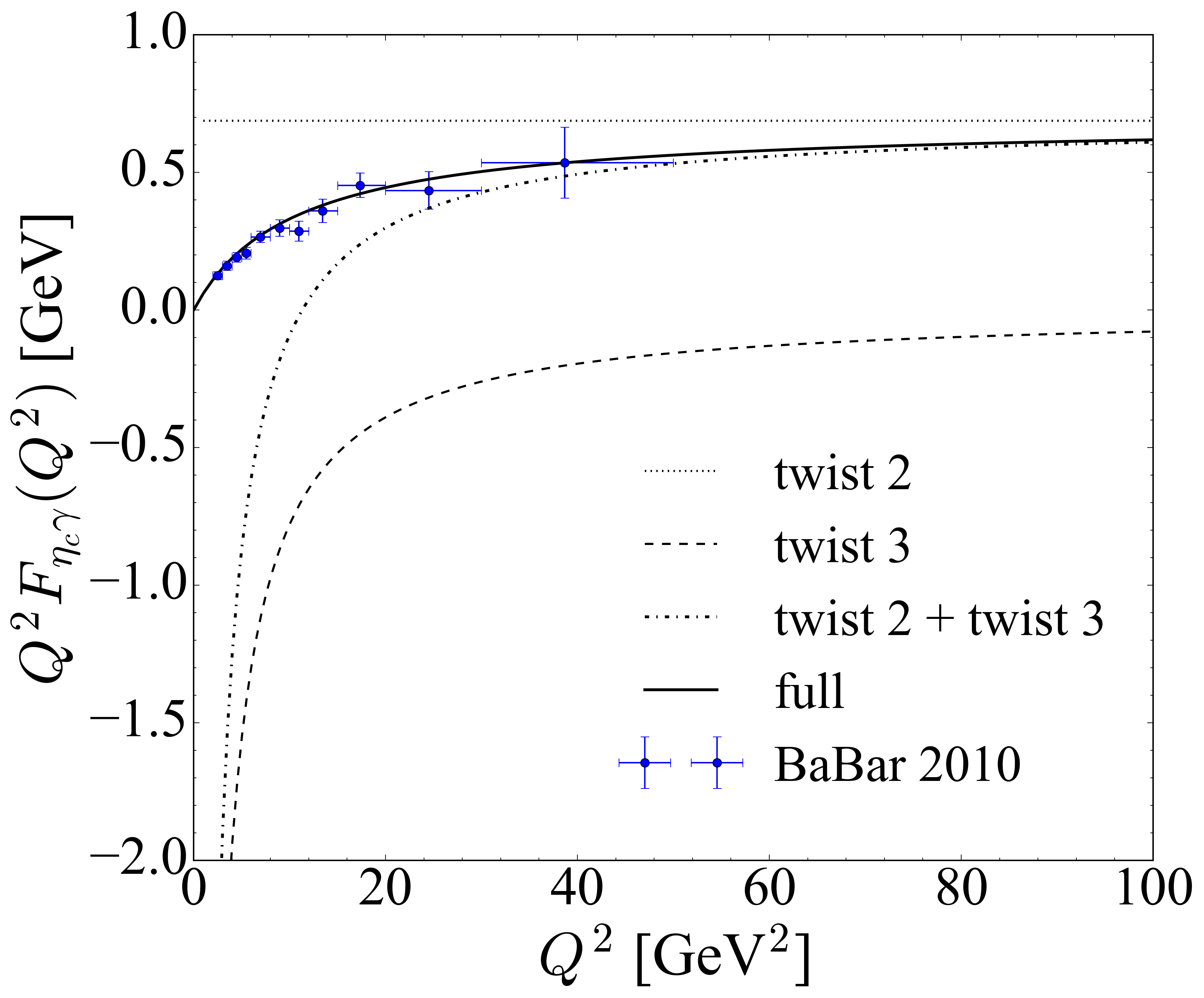}
\begin{center}
\includegraphics[width=0.37\columnwidth,clip=]{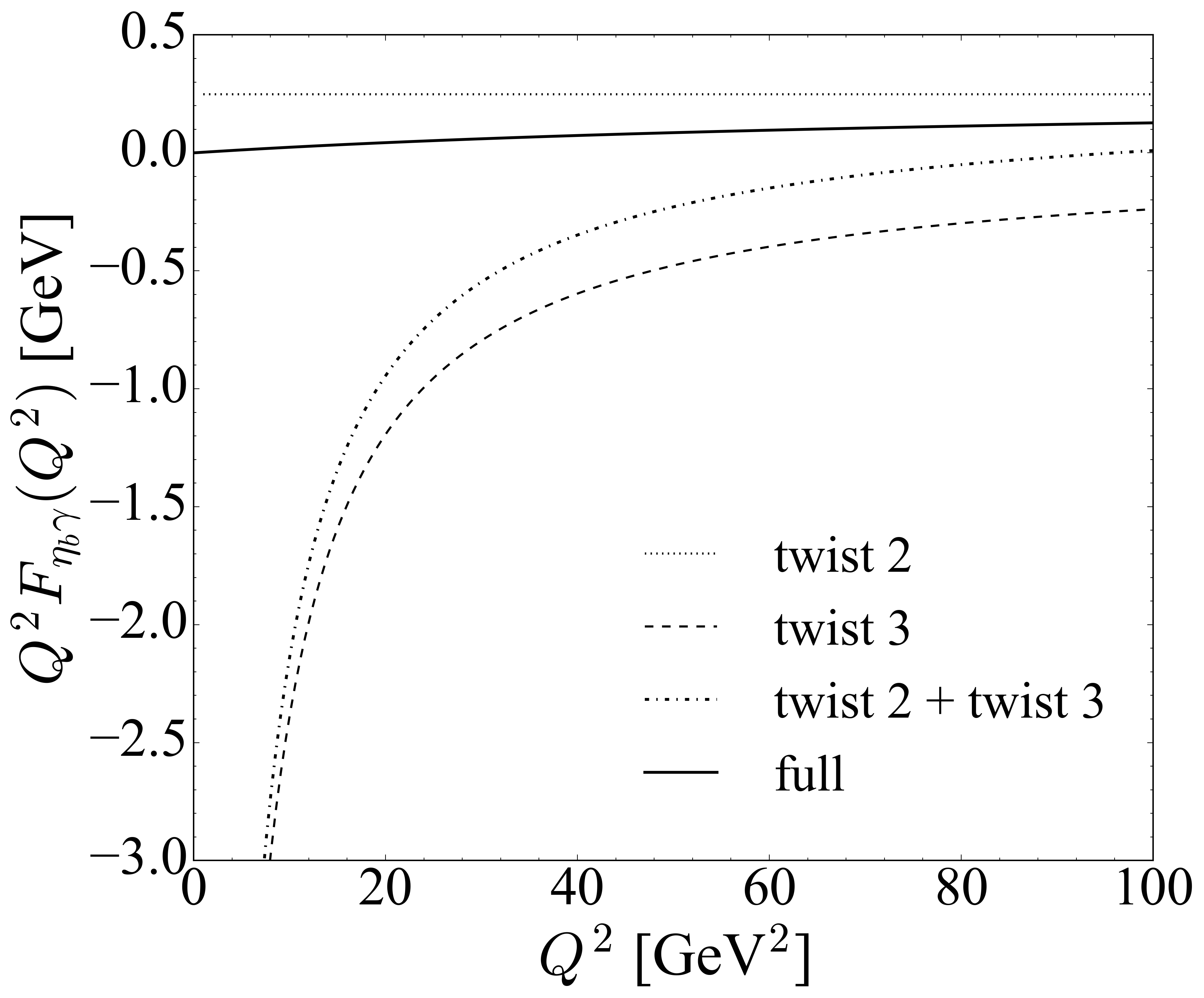}
\end{center}
\caption{The contributions of the leading- and higher-twist DAs to the transition form factors $Q^2F_{(\pi,\eta_c,\eta_b)\gamma}(Q^2)$ in the
spacelike  momentum transfer region ($0<Q^2<100$ GeV$^2$). 
}
%\end{center}
\end{figure*}

In Fig. 3, we show the normalized $\eta_c\to\gamma\gamma^*$ transition form factor $F_{\eta_c\gamma}(Q^2)/F_{\eta_c\gamma}(0)$ obtained from the Set II
for both timelike ($q^2=-Q^2>0$)
spacelike ($q^2=-Q^2<0$) momentum transfer regions up to $|Q^2|=70$ GeV$^2$ and compare them with
the available experimental data~\cite{Lees:2010de}  for the spacelike region as well as 
the results obtained from the dispersion relation (DR).
The dotted, dashed and solid lines in Fig.~3 represent our LFQM predictions
of  ${\rm Re}\;[F_{\eta_c\gamma}(q^2)/F_{\eta_c\gamma}(0)]$, ${\rm Im}\;[F_{\eta_c\gamma}(q^2)/F_{\eta_c\gamma}(0)]$ 
and $|F_{\eta_c\gamma}(q^2)/F_{\eta_c\gamma}(0)|$, respectively.  
We note that the spacelike  region can
be easily obtained by analytically continuing the momentum transfer $q^2\to -q^2$ in the integrand of Eq.~(\ref{Eq3}).
As one can see from Fig.~3,
our result for the spacelike $Q^2$ region shows a good agreement with the data.
For the analysis of  timelike form factor near resonance region in Fig.~3,
the maximum value of  $F_{\eta_c\gamma}(q^2)$  occurs  at  $q^2\simeq 4m^2_c$ due to the virtual photon wave function term
$1/(M^2_0 -q^2)$ in Eq.~(\ref{Eq3}). The imaginary part of the form factor also starts to appear at  $q^2 = 4m^2_c$.
As a consistency check of our LFQM calculations for the timelike region, 
we also include the real (imaginary) part of the form factor  obtained 
from the DR (denoted by $+(\times)$ data points) given by Eq. ~(\ref{ReF}). 
As one can see, our direct results for  the real and  imaginary parts are in perfect agreement with the results obtained from the DR. 
This assures the validity of our numerical calculation in the timelike region.

In Fig.~4, we show the normalized TFFs
$F_{\eta_c\gamma}(Q^2)/F_{\eta_c\gamma}(0)$ (left panel) for the spacelike ($q^2=-Q^2<0$) momentum transfer region
up to $Q^2=100$ GeV$^2$
and $|Q^2 F_{\eta_c\gamma}(Q^2)|$ (right panel)
for both timelike ($q^2>0$) and spacelike  momentum transfer regions ($-500 \leq Q^2\leq 500$ GeV$^2$)
and compare them with the available experimental data~\cite{Lees:2010de}
for the spacelike region. The dashed and solid lines  represent our results obtained from
the Set I and II, respectively. We note that the spacelike  region can
be easily obtained by analytically continuing the momentum transfer $q^2\to Q^2(=-q^2)$ in the integrand of Eq.~(\ref{Eq3}).
Our results from the Set II are in good agreement with the available data not only for 
the normalized TFF $F_{\eta_c\gamma}(Q^2)/F_{\eta_c\gamma}(0)$
but also for the form factor  $F_{\eta_c\gamma}(0)$ at $Q^2=0$.
We note that our LFQM result for $|Q^2 F_{\eta_c\gamma}(Q^2)|$  shows the asymptotic behavior for high $|Q^2|$ values, but
the result in the spacelike region reaches the asymptotic value faster than that in the timelike region.

In Fig.~5, we show the normalized TFFs
$F_{\eta_b\gamma}(Q^2)/F_{\eta_b\gamma}(0)$ (left panel) for the spacelike  momentum transfer region
up to $Q^2=100$ GeV$^2$ and $|Q^2 F_{\eta_b\gamma}(Q^2)|$ (right panel)
for both timelike and spacelike  momentum transfer regions ($-500 \leq Q^2\leq 500$ GeV$^2$).  
The line codes are same as in Fig.~4.
While the qualitative behavior of the
$F_{\eta_b\gamma}$ is the same as that of $F_{\eta_c\gamma}$, their quantitative behaviors such as the slope of the form factor at $Q^2=0$ 
are quite different due to the $b$ quark being much heavier than the $c$ quark.
Our LFQM result for $|Q^2 F_{\eta_b\gamma}(Q^2)|$  shows the asymptotic behavior for high $|Q^2|$ values, but again
the result in the spacelike region reaches the asymptotic value faster than that in the timelike region. 

In Fig.~6, we show the contributions of the leading- and higher-twist DAs to the transition form factors $Q^2F_{(\pi,\eta_c,\eta_b)\gamma}(Q^2)$ in the
spacelike  momentum transfer region ($0<Q^2<100$ GeV$^2$). The dotted, dashed, and dot-dashed lines represent the contributions from the twist-2 DAs
$\phi_{2;M}(x)$, the twist-3 DAs $\phi_{3;M}(x)$, and the sum of the twist-2 and twist-3 DAs (see Eq.~(\ref{Q2F})), respectively.  The solid line represents
the full results of $Q^2F_{(\pi,\eta_c,\eta_b)\gamma}(Q^2)$ given by Eq.~(\ref{Eq3}). The results for the heavy quarkonia are obtained from the Set II parameters.
As one can see, most of the contributions to $Q^2F_{\pi\gamma}(Q^2)$ for $Q^2 \geq 10$ GeV$^2$ come from the pion DAs up to twist-3 and the contributions from
the twist-4  DAs and above are negligible for $Q^2 \geq 10$ GeV$^2$ region. On the other hand, for the $Q^2F_{\eta_c\gamma}(Q^2)$ case, the contributions from
the twist-2 and twist-3 DAs are dominant only after $Q^2> 60$ GeV$^2$. This indicates that the higher twist contributions beyond the twist-3 contribution 
are not negligible to fit the currently
available experimental data for $Q^2F_{\eta_c\gamma}(Q^2)$. For the $Q^2F_{\eta_b\gamma}(Q^2)$ case, our LFQM shows the necessity of the higher twist contributions 
beyond the twist-3 contribution even for $Q^2>100$ GeV$^2$.

\section{Conclusions}
\label{sec:V}
We studied the $(\eta_c,\eta_b)\to\gamma^*\gamma$ transitions for the entire kinematic regions analyzing both spacelike and timelike TFFs in 
our LFQM.  
Especially, the calculations of $F_{\eta_c\gamma}$ and $F_{\eta_b\gamma}$ have been performed by our newly developed method
using the $q^+\neq 0$ frame with $q^+=P^+$~\cite{Choi:2017zxn}, which
is found to be most effective for the analysis of the timelike region due to the absence of mixing between the internal transverse momentum 
and the external virtual photon momentum. This leads to the very simple pole structure $1/(q^2 - M^2_0)$ in the form factor, which  
not only  leads to the emergence of the imaginary part of the form factor starting at $q^2=4 m^2_Q(Q=c,b)$ but also
provides a straightforward systematic twist expansion of TFFs.
We obtained the twist 2 and 3 TMDAs as well as the corresponding twist 2 and 3 DAs in this work using our LFQM framework.
As a consistency check for our numerical calculations in timelike region, we have confirmed that our direct LFQM results 
of $F_{\eta_c(\eta_b)\gamma}(Q^2)$ are in excellent agreement with those obtained from the dispersion
relations. 

In our numerical calculation of the normalized TFF $F_{\eta_c\gamma}(Q^2)/F_{\eta_c\gamma}(0)$ and the decay width
$\Gamma_{\eta_c\to\gamma\gamma}$,
our LFQM results from $m_c=1.3$ GeV are more consistent with the data~\cite{Lees:2010de,PDG2018} than the results from $m_c=1.8$ GeV.
Compared to the light pseudoscalar meson TFFs such as $(\pi^0,\eta,\eta')\to\gamma\gamma^*$ transitions analyzed in~\cite{Choi:2017zxn},
the completely symmetric asymptotic behaviors for the heavy $|Q^2F_{(\eta_c,\eta_b)\gamma}(Q^2)|$ TFFs independent of the timelike and spacelike
regions are not reached within a few hundred GeV$^2$  values of $|Q^2|$.
%but delayed until $|Q^2|\sim 10^4$ GeV$^2$ for the $F_{\eta_c\gamma}$ case and even higher $|Q^2|$ values for the $F_{\eta_b\gamma}$ case. 
This 
may be due to the resonance structure occurring at large $q^2 \simeq 4m^2_Q(Q=c,b)$ in the timelike region. 
More elaborate LFQM calculation deserves further study including more trial wave functions such as 2$S$ state and even higher excited radial state harmonic oscillator wave functions.

%\section*{Acknowledgment}
\acknowledgments
H.-Y. Ryu was supported by the NRF grant funded by
the Korea government(MSIP) (No. 2015R1A2A2A01004238).
H.-M. Choi was supported by the National Research Foundation of Korea (NRF)
(Grant No. NRF-2017R1D1A1B03033129). 
C.-R. Ji was supported in part by the US Department of Energy
(Grant No. DE-FG02-03ER41260).

\end{document}